\documentclass[useAMS,usenatbib]{mn2e} 
\usepackage{epsfig,color}
\usepackage{psfrag,graphicx,amsmath,multicol}
\usepackage[dvipsnames]{xcolor}
\definecolor{webgreen}{rgb}{0,.5,0}
\definecolor{webbrown}{rgb}{.6,0,0}

\newcommand{\comments}[1]{}
\newcommand       \be           {\begin{equation}}
\newcommand       \ee           {\end{equation}}
\newcommand       \ba           {\begin{eqnarray}}
\newcommand       \ea           {\end{eqnarray}}

\newcommand       \aj          {AJ}
\newcommand       \apj          {ApJ}
\newcommand       \apjl         {ApJL}
\newcommand       \apjs         {ApJS}

\newcommand       \nat          {Nature}
\newcommand       \mnras        {MNRAS}
\newcommand       \araa         {Ann. Rev. Astr. Astr.}

\def\msun{\rm \ M_\odot}
\def\mpy{\rm \ M_\odot~{\rm yr^{-1}}}

\def\lesssim{\mathrel{\hbox{\rlap{\hbox{\lower4pt\hbox{$\sim$}}}\hbox{$<$}}}}
\def\gtrsim{\mathrel{\hbox{\rlap{\hbox{\lower4pt\hbox{$\sim$}}}\hbox{$>$}}}}

\defcitealias{mcc11}{Paper~I}
\defcitealias{sha11}{Paper~II}
\defcitealias{voi02}{V02}

\title[$L_X-T_X$ Relation \& Missing Baryons]
{On the Structure of Hot Gas in Halos: Implications for the $L_X-T_X$ Relation \& Missing Baryons}  

\author[P.\ Sharma, M.\ McCourt, I.~J.\ Parrish, E.\ Quataert]
{Prateek Sharma$^\dag$, Michael McCourt$^\ddag$, \& Ian J. Parrish$^\ddag$, Eliot Quataert$^\ddag$ \\
$^\dag$Department of Physics and Joint Astronomy Program, Indian Institute of Science, Bangalore, India 560012 (prateek@physics.iisc.ernet.in) \\
$^\ddag$Astronomy Department and Theoretical Astrophysics Center,
B-20 Hearst Field Annex 3411, University of California, Berkeley, CA 94720, USA}

\begin{document}

\pagerange{\pageref{firstpage}--\pageref{lastpage}} \pubyear{2012}
\maketitle
\label{firstpage}

\begin{abstract}
We present one-dimensional models of the hot gas in dark-matter halos, which both
predict the existence of cool cores and explain their structure.  Our models
are directly applicable to semi-analytic models (SAMs) of galaxy formation.
We have previously argued that filaments of cold ($\sim 10^4$\,K) gas condense
out of the intracluster medium (ICM) in hydrostatic and thermal equilibrium when the ratio of the thermal instability timescale to
the free-fall time $t_{\mathrm{TI}}/t_{\mathrm{ff}}$ falls below 5--10.  This
criterion corresponds to an upper limit on the density of the ICM and motivates a model in which a density core forms wherever
$t_{\mathrm{TI}}/t_{\mathrm{ff}}~\lesssim~10$.  Consistent with observations
and numerical simulations, this model predicts larger and more tenuous cores
for lower-mass halos---while the core density in a cluster may be as large
as $\sim 0.1$\,cm\textsuperscript{-3}, the core density in the Galactic halo
should not exceed $\sim 10^{-4}$\,cm\textsuperscript{-3}.  { }{We can also explain the large densities in smaller mass halos (galactic `coronae') if we include the contribution of the central galaxy to the gravitational potential.} Our models produce a favorable match to
 the observational X-ray luminosity-temperature ($L_{\mathrm{X}}-T_{\mathrm{X}}$) relation. 
 For halo masses $\lesssim 10^{13} \msun$ the core size approaches the virial radius. Thus, most of the baryons in such halos cannot be in the hot ICM, but either in the form of stars
 or in the form of hot gas beyond the virial radius. Because of the smaller mass in the ICM and much
 larger mass available for star formation, the majority of the baryons in low mass halos ($\lesssim 10^{13} \msun$) can be expelled beyond the virial radius 
 due to supernova feedback. This can account for the baryons `missing' from low mass halos, such as the Galactic halo.
\end{abstract}


\section{Introduction}

In the currently favored $\Lambda$CDM cosmology, massive halos form due to mergers of smaller dark matter halos. The overall structure of the halos is simply an outcome of gravitational interactions between dissipationless dark matter particles (e.g., \citealt{spr05}), and once the dark matter halo relaxes the density takes a simple self-similar form (\citealt{nav97}). While the evolution of dark matter halos is simple, at least in principle, the evolution of baryons residing in these halos is much more complicated. In addition to gravity, the baryons are affected by complex, difficult to model processes such as radiative cooling, feedback heating, and turbulence.

Observationally it is well known that the distribution of baryons in the form of hot gas and condensed phases (stars, molecular and atomic gas) is a sensitive function of  halo mass. Baryons in the most massive halos (clusters and groups) are mainly in the form of the hot, X-ray emitting plasma. The baryon fraction in the hot phase decreases with a decreasing halo mass, but the fraction in stars increases to keep the total baryon fraction roughly constant for halos $\gtrsim 10^{13} \msun$ (\citealt{gon07,gio09}). While massive halos seem baryonically `closed', smaller halos, including our own Milky Way, have lost most of their baryons (\citealt{dai10,mcg10}).  Observationally, the halo mass seems to be the primary variable determining the halo baryon and stellar fractions, { }{irrespective of AGN activity or star formation} (\citealt{and10,mcg10}). This remarkable correlation between the halo mass and stellar/baryonic mass provides an important clue to understanding galaxy formation. A related observational result is that the smaller halos have lower X-ray luminosities ($L_X$) than self-similar models which assume a similar gas density/temperature profile for all halos. In this paper we present models  which explain these observations in terms of the interplay of cooling, feedback heating, and thermal instability in halos.

Cooling and feedback play a fundamental role in galaxy formation. Both cooling and star formation are efficient if the cooling time at the virial radius is shorter than the dynamical/free-fall time (e.g., \citealt{ree77,whi78}). Even when the cooling time is longer than the free-fall time, it can be shorter than the halo age ($\sim$ Hubble time). Thus, in absence of additional heating, the inner radii of halos affected by cooling are expected to form massive cooling flows (e.g., \citealt{fab94}). However, both the absence of soft X-ray lines (e.g., \citealt{pet03}) and the lack of significant star formation (e.g., \citealt{ode08}), indicate that cooling in cluster cores is largely suppressed. There is growing evidence that active galactic nuclei (AGN) jets, bubbles, and cavities (e.g., \citealt{bir04}) blown by the central supermassive black holes (SMBHs) in cluster cores play a crucial role in suppressing cooling flows (see \citealt{mcn07} for a review). Supernova feedback in lower mass halos (e.g., \citealt{efs00}) appears sufficient to prevent runaway cooling of the hot gas. Moreover, feedback heating is globally stable because the condensation of cold gas leads to enhanced heating (e.g., see \citealt{piz05}). Understanding the hot gas thermodynamics is essential to explain the non-self-similarity of baryons in halos, as we argue below.

Motivated by observations which suggest balance between heating and cooling, in a series of papers (\citealt{mcc11}, hereafter \citetalias{mcc11}; \citealt{sha11}, hereafter \citetalias{sha11}) we studied the role of local thermal instability in governing the structure of the pressure-supported hot gas ($\sim 10^6-10^8$ K) for  virialized halos in global thermal balance. Our idealized simulations, which impose a balance between radiative cooling and heating, predict that cold filaments condense out of the hot phase if, and only if, the  thermal instability timescale ($t_{\rm TI}$)  is shorter than approximately ten times the free-fall time ($t_{\rm ff}$).\footnote{Cold gas of any kind is referred to as filaments in this paper; numerical simulations and observations suggest that the cold gas is filamentary. For a heating rate per unit volume which is independent of density, the thermal instability timescale is roughly equal to the cooling time; for definitions of various timescales see Eqs. 8-10 in \citetalias{sha11}.} This criterion gives an upper limit on the hot gas density in cores of different halo masses. For gas densities larger than this, cold filaments condense out of the hot phase. The cold overdense filaments are heavier than the surrounding hot gas, and thus fall toward the center and power AGN/supernovae. Thus, even if feedback heating can balance cooling in the cluster core, the physics of local thermal instability gives a rough upper limit on the  density of the hot gas. The end result is that the core density is reduced until the criterion $t_{\rm TI}/t_{\rm ff} \gtrsim 10$ is satisfied for the hot gas. 

{ }{Our cores with $t_{\rm TI}/t_{\rm ff} \sim 10$ agree with the observations of cool core clusters.} Our results are robust as long as the timescale for secular cooling/heating of the ICM is longer than the radiative cooling time; i.e., if the halo is in rough thermal balance. The reader should consult \citetalias{mcc11} \& \citetalias{sha11} for more details. Recent, more realistic cluster simulations with AGN jet feedback \citep[][]{gas12} roughly agree with the findings of our idealized simulations; { }{in particular, extended multiphase gas condenses out of the ICM and leads to enhanced feedback when our $t_{\rm TI}/t_{\rm ff} \lesssim 10$ criterion is satisfied. However, as expected, there are differences in detail; e.g., the presence of angular momentum supported cold torus at small radii.}

Based on our $t_{\rm TI}/t_{\rm ff}$ criterion for the formation of multiphase gas, it is straightforward to build one-dimensional hydrostatic models for the structure of the hot gas in virialized halos (see \citealt{voi02} for a similar approach). We assume a universal gas density profile at large radii unaffected by cooling/feedback, and we build hydrostatic profiles inward until the radius where $t_{\rm TI}/t_{\rm ff}$ is smaller than the threshold for forming cold filaments. An isentropic core is introduced within that radius,  as suggested by our numerical simulations and by observations \citep[][]{cav09}. We explore our models by changing several parameters, the outer gas density profile, $t_{\rm TI}/t_{\rm ff}$ threshold, a constant $t_{\rm TI}/t_{\rm ff}$ core, { }{inclusion of stellar gravitational potential}, etc. Our results show that lower mass halos have larger but lower density cores because of shorter cooling times. This is consistent with group observations (\citealt{sun12}) and with numerical simulations (see Fig. 9 in \citetalias{sha11}). The X-ray luminosity-temperature ($L_X-T_X$) relation derived from our models is also consistent with observations. Moreover, we do not expect much redshift evolution of the hot gas in virialized halos because the $t_{\rm TI}/t_{\rm ff}$ criterion, and hence the core density, is essentially independent of the redshift. 

We estimate the feedback efficiency required to balance radiative cooling ($L_X$) in halos with different masses. We find that large feedback efficiencies ($0.001-0.01$) are required for massive clusters. However, for lower halo masses the core is extended and has a lower density. Assuming that most of the baryons  that are absent from the ICM are channeled into star formation and accretion, implies that a smaller feedback efficiency is required for lower mass halos such as groups. For halos of even smaller mass ($\lesssim 10^{13} \msun$) the required feedback efficiency is extremely small ($\lesssim 10^{-7}$), $10-100$ times  smaller than what is available due to supernova feedback. This means that most of the baryons in such halos will be overheated and will be pushed beyond the virial radius. Therefore, for lower mass halos, such as our own Galaxy, most baryons which should have been accreted in the absence of feedback, are `missing'. These `missing' baryons are in the extended corona which is prone to forming multiphase gas required to explain quasar absorption lines observed close to galaxies (e.g., \citealt{che10}).

This paper is organized as follows. We describe our one-dimensional models in Section 2. In Section 3 we derive the X-ray luminosity-temperature ($L_X-T_X$) relation from our models and compare it with observations. Section 3 also presents the implications of our models for the `missing' baryons problem. We conclude in Section 4 with astrophysical implications of our models.

\section{One-dimensional Models for Hot Gas in Halos}

We closely follow the comprehensive paper by \citealt{voi02} (hereafter \citetalias{voi02}; see also \citealt{fuk06}) to build our one-dimensional models. Nevertheless, there are significant differences. The key difference is that they apply an entropy floor within the radius where the cooling time is shorter than the cluster age, but we apply an entropy floor (we also try a floor on $t_{\rm TI}/t_{\rm ff}$, as described later) where our $t_{\rm TI}/t_{\rm ff}\lesssim 10$ threshold is satisfied. The physical pictures corresponding to the two criteria are different. The \citetalias{voi02} model posits that the X-ray gas with a cooling time shorter than the cluster age either cools to the cold phase or is heated by feedback such that the cooling time is longer than the cluster age. Theoretically, however, steady-state cooling flows maintain cooling times much shorter than the Hubble time. The cooling scenario is further disfavored as it suffers from the overcooling problem. The core entropy obtained by equating the cooling time to the cluster age ($\sim 5$ Gyr; e.g., Fig. 1 in \citealt{voi01}) is much larger than what is observed in cool core clusters (e.g., \citealt{cav09}), and there are also non-cool core clusters with cooling times longer than the cluster age.  { }{We note, of course, that the \citetalias{voi02} model seems to satisfactorily describe the non-cool core cluster density profiles and the resulting $L_X-T_X$ relation.}

In contrast to \citetalias{voi02}, our threshold is based on the physics of local thermal instability in cluster halos. The central assumption of our models is that heating and cooling roughly balance globally and regulate the core to an approximate thermal equilibrium. While we do not understand the details of how the ICM is heated, the assumption of rough thermal balance is robust and observationally motivated for cool core clusters. The  lack of massive cooling flows suggests that the heating rate should be at least as large as the radiative cooling rate. Likewise, heating cannot be too dominant; otherwise the cold gas would be totally evaporated (see Fig. 13 in \citealt{sha10}). Thus, cluster cores are expected to be in rough thermal balance over a few cooling times.

 The physics of local thermal instability in the ICM (we refer to the hot gas in all virialized halos, even groups and individual galaxies, as the ICM) in thermal balance introduces a new parameter $t_{\rm TI}/t_{\rm ff}$ which is absent in earlier one-dimensional models.  The thermal instability occurs in two regimes depending on $t_{\rm TI}/t_{\rm ff}$. If this ratio is smaller than a critical threshold ($\sim 10$) then local thermal instability leads to multiphase gas; an overdense blob cools to the stable temperature and falls toward the cluster center on $\sim$ free-fall timescale. For a larger $t_{\rm TI}/t_{\rm ff}$, the overdense blob responds to gravity as it is cooling. Shear generated between the infalling overdense blob and the background hot gas is able to mix it before it can cool to the thermally stable phase. Thus, no multiphase gas is expected if $t_{\rm TI}/t_{\rm ff} \gtrsim 10$. { }{We argue in \citetalias{sha11} that, since extended multiphase gas leads to very high accretion rates, cooling and feedback in cool cores should self-regulate to the critical threshold where $t_{\rm TI} \sim 10 t_{\rm ff}$}. Our $t_{\rm TI}/t_{\rm ff}$ criterion results in a realistic core entropy for cool core clusters and performs well as a predictor of extended cold gas in cluster cores (e.g., see Fig. 12 in \citealt{mcc11}). The physical basis of our models and good quantitative match with observations give us  confidence in the predictions of our one-dimensional models. 
 
 { }{We point out that, though our criterion on the ratio $t_{\rm TI}/t_{\rm ff}$ is very different from the criterion on the cooling time in \citetalias{voi02} for group and cluster masses, the two models yield similar results for Galactic mass halos.  This is because we find that the core size in the Galactic halo is $\sim$ the virial radius, where the dynamical time is $\sim 1/10^{\rm th}$ of the Hubble time.  Thus, for $10^{12}$ solar mass halos, our criterion that $t_{\rm TI}/t_{\rm ff} \sim 10$ is equivalent to the assumption that the cooling time roughly equals the halo age.  \citet{mal04} explored models of the Galactic halo in which the cooling time at the virial radius is matched to the halo age; our results in this mass range agree closely with theirs. The long cooling time at the virial radius also signals the breakdown of our model, however: if the cooling time is comparable to the age of the system, our assumption of thermal equilibrium need not hold.  Thus, our results for halo masses below $\sim 6 \times 10^{12} \msun$ should be viewed more as an extrapolation from higher masses than as solid quantitative predictions.}

\subsection{Recipe for Constructing One-dimensional Models}

We construct one dimensional gas profiles such that the ICM is in hydrostatic equilibrium and everywhere satisfies our criterion $t_{\rm TI}/t_{\rm ff} \gtrsim 10$. In determining hydrostatic equilibrium we neglect gravity due to gas and stars for most models, and assume a fixed NFW dark matter potential \citep[][]{nav97},  with the concentration parameter ($c \equiv r_{200}/r_s$, where $r_{200}$ is the virial radius within which the average density is 200 times the critical density and $r_s$ is the scale radius) adjusted according to the halo mass. The relation between the concentration parameter ($c$) and $M_{200}$ (mass within $r_{200}$) is based on \citetalias{voi02} and references therein (see Table \ref{tab:tab4}). Our results are only weakly sensitive to $c$.

Following \citetalias{voi02}, we construct an `unmodified' profile in which the gas density is specified and other thermodynamic variables are calculated using hydrostatic equilibrium. The `unmodified' density profile agrees with the observed ICM density at large radii and has a baryon fraction equal to the cosmic value. Thus, the total gas mass in the `unmodified' halo equals $f_bM_{200}$, where $f_b=0.17$ is the universal baryon fraction. Although observations suggest that smaller halos are baryon poor,  for simplicity we use a fixed baryon fraction for all `unmodified' halos. As we show in Section \ref{sec:BB}, the final baryon fraction is something we can calculate using our models. Our results are only weakly dependent on the assumed initial value of $f_b$. 

The effect of cooling and heating on the ICM is incorporated by altering the initial profile so that it satisfies $t_{\rm TI}/t_{\rm ff} \gtrsim 10$ everywhere. This process amounts to introducing a low density core in the center of the cluster. Since this prescription is not unique, we try two different core profiles: an isentropic ($T/\rho^{2/3} =$ constant) core; and  a `core' with a fixed $t_{\rm TI}/t_{\rm ff}$. Each constraint, along with hydrostatic equilibrium, then uniquely specifies the structure of the core. In each case, the atmosphere outside the core is unmodified. 

Observations suggest that an isentropic core is a good approximation for the inner radii of the ICM, for both cool core and non-cool core clusters (see \citealt{cav09}). A `core' with $t_{\rm TI}/t_{\rm ff}$ equal to the critical value for forming cold filaments represents an upper limit on the density of the hot phase at all radii within the core. If the density exceeds this limit, a large mass of cold gas will condense out of the hot phase, leaving behind a core with $t_{\rm TI}/t_{\rm ff} \sim 10$ (\citetalias{sha11}).

 Since the gas density profile, even in cluster outskirts, is not universal (due to mergers, accretion history, cosmological filaments, etc.; e.g., \citealt{cro08}), we try different `unmodified' gas density profiles of a generalized NFW form
\be
\label{eq:dpro}
\rho = \frac{N_c}{ r/r_s (1+ r/r_s)^{(-s-1)}},
\ee
where $N_c$ is a normalization constant such that 
$$\int_0^{r_{200}} 4\pi r^2 \rho dr = f_b M_{200},$$
 and $s$ is the asymptotic ($r \rightarrow \infty$) gas density slope $d\ln \rho/d\ln r$ ($s=-3$ for an NFW density profile). 
 
The last parameter needed to specify our models is a boundary condition on one of our thermodynamic variables. We choose the pressure at $r_{200}$ as the outer boundary condition. The outer pressure is specified as $p_{200}=k_s \langle n \rangle k_B T_{200}$, where $k_s$ is a constant, $\langle n \rangle \equiv 3 f_b M_{200}/4\pi \mu m_p r_{200}^3$  ($\mu$ is the mean molecular weight, $m_p$ is the proton mass)  is the average gas number density within $r_{200}$, and $k_B T_{200} \equiv GM_{200} \mu m_p/2r_{200}$ is the virial temperature at $r_{200}$. The constant $k_s$ is smaller than unity since the number density at $r_{200}$ is smaller than the average density $\langle n \rangle$; moreover, $k_s$ should be smaller for a steeper initial gas density profile. We choose $k_s=0.6,~0.32,~0.15$ for $s=-1.75,~-2.25,~-3$ respectively. These values ensure that a power-law entropy profile is obtained at large radii for  `unmodified' profiles in hydrostatic equilibrium. The results are quite insensitive to the exact value of $k_s$.
 
We calculate our entropy/temperature profiles starting at $r_{200}$ and moving in by imposing hydrostatic equilibrium and using the ICM density given by Eq. \ref{eq:dpro}. We calculate the $t_{\rm TI}/t_{\rm ff}$ ratio (see Eqs. 8-10 in \citetalias{sha11} for definitions) at each radius. We use the fit in \citet{toz01} (Eq. B4, Table 3) based on \citet{sut93} for the cooling function corresponding to one-third of the solar metallicity (see Fig. 1 in \citealt{sha10}). If $t_{\rm TI}/t_{\rm ff}$ at some radius becomes smaller than the critical value (chosen to be 10 for most of our models) we introduce a `core' within that radius. We use the same critical value of $t_{\rm TI}/t_{\rm ff}$ irrespective of the halo mass. This assumption is justified if halos are in rough thermal balance and the microphysical heating mechanism is similar for all of them.

 For smaller halos ($\lesssim 10^{13} \msun$), $t_{\rm TI}$ at the virial radius can be shorter than $10 t_{\rm ff}$ because of efficient line cooling. In these cases, we adjust the density and the temperature at the outer boundary such that $t_{\rm TI}/t_{\rm ff}$ there equals the critical value and the pressure equals the imposed value at the outer boundary. Since the results of our one-dimensional models in such cases depend on the outer boundary conditions (e.g., whether we keep the outer pressure/temperature/density fixed), the structure of the hot plasma in lower mass halos can only be studied in detail by cosmological numerical simulations. However, most of our results stem from the trend of core size with halo mass, which we expect to hold even for the lower mass halos.  It is important to note that the limit on $t_{\rm TI}/t_{\rm ff}$  ($\sim 10$) is a lower limit { }{which should apply to cool core halos}; the minimum $t_{\rm TI}/t_{\rm ff}$ for gas can be larger than $5-10$ in non-cool core systems because of overheating by feedback and/or major mergers. 

Here we  briefly summarize the various steps involved in constructing our one-dimensional models:
\begin{enumerate}
\item Assume an `unmodified' gas density profile of a generalized NFW form (Eq. \ref{eq:dpro}), such that the total gas mass is $f_b$ times the halo mass.
\item Specify the pressure at $r_{200}$ as an outer boundary condition and construct `unmodified' profiles by assuming the gas density profile given by Eq. \ref{eq:dpro} and imposing hydrostatic equilibrium in an NFW potential ({ }{we also include gravity due to the central galaxy in some models}); choose $p_{200}$ such that the entropy profile close to the virial radius is a power-law.
\item Calculate $t_{\rm TI}/t_{\rm ff}$ at each radius moving inside from the outer boundary; introduce a `core' with a constant entropy or a constant $t_{\rm TI}/t_{\rm ff}$ within the radius where $t_{\rm TI}/t_{\rm ff}$ becomes smaller than the critical value for cold gas condensation.
\item Impose hydrostatic equilibrium within the `core' by assuming  a constant entropy (or $t_{\rm TI}/t_{\rm ff}$) and obtain the `modified' profile.
\end{enumerate}

\section{Results}

\begin{figure}
\centering
\includegraphics[scale=0.42]{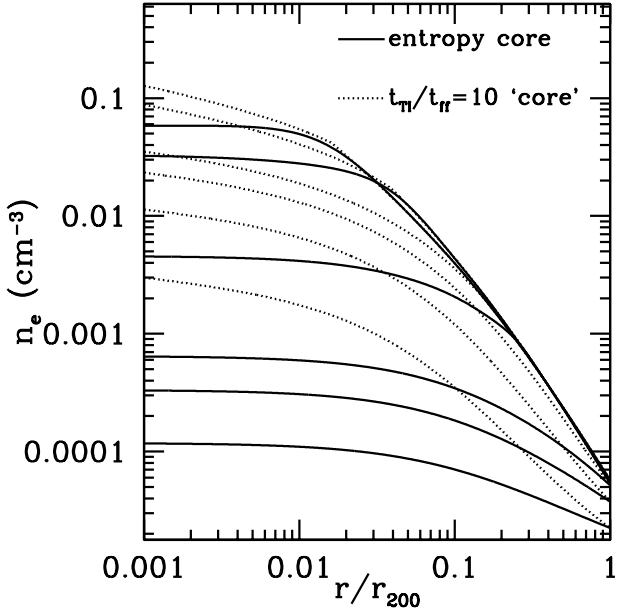}
\caption{The electron number density ($n_e$) as a function of radius (scaled by $r_{200}$) for different halo masses; the asymptotic gas density slope $s=-2.25$ for these models. The profiles for an entropy core with $t_{\rm TI}/t_{\rm ff} \geq 10$ are shown by the solid lines and the profiles for a fixed $t_{\rm TI}/t_{\rm ff} = 10$ `core' are shown by the dotted lines. The halo masses, in the order of decreasing core density, are $3\times10^{14}$, $6\times 10^{13}$, $10^{13}$, $6\times 10^{12},~3\times 10^{12},~{\rm and}~10^{12}~\msun$.
\label{fig:nvsr}}
\end{figure}

Table \ref{tab:tab4} contains our models for halos ranging from the Galactic mass ($10^{12} \msun$) to most massive clusters ($3 \times 10^{15} \msun$). This table lists various quantities calculated from our constant entropy and constant $t_{\rm TI}/t_{\rm ff}$ models with the critical $t_{\rm TI}/t_{\rm ff}=10$. The table lists the central electron density, the central entropy, the X-ray luminosity-weighted temperature ($T_X$), and the X-ray luminosity ($L_X$). Table \ref{tab:tab4}  also lists the mass dropout factor $f_d$, the fraction of the baryonic mass which is removed from the hot ICM when we `modify' the core. This mass can either go into fueling star formation and black hole growth, or can be pushed outside $r_{200}$ because of strong feedback heating (in this sense the term `dropout factor' is misleading because it includes hot gas beyond the virial radius). In addition, the table lists an estimate of the mass dropout rate $\dot{M}_d$, the ratio of the mass absent from the hot ICM and the typical halo age $t_{\rm age}$. The halo age is assumed to be 5 Gyr for all halos; merger rate, and hence $t_{\rm age}$, is very weakly dependent on the halo mass; e.g., see \citealt{fak10}. From $\dot{M}_d$, we estimate the required feedback heating as the ratio of  the X-ray luminosity and $\dot{M}_dc^2$, $\epsilon_{\rm req} \equiv L_X/\dot{M}_dc^2$. 
 
 \begin{figure}
\centering
\includegraphics[scale=0.42]{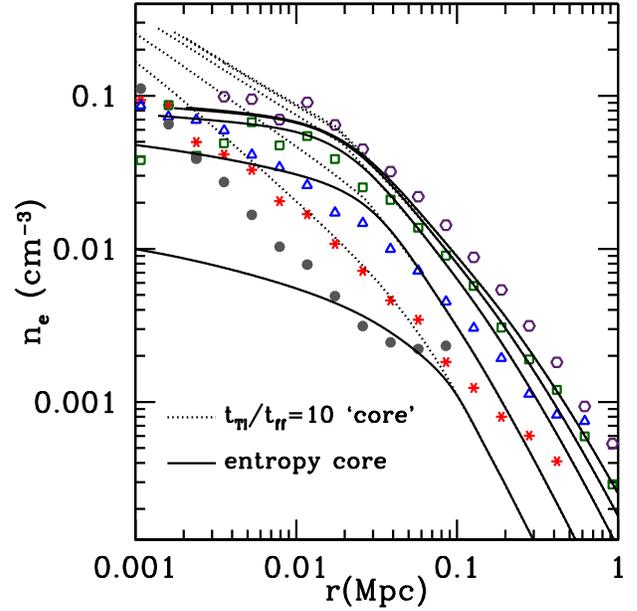}
\caption{ { }{Electron number density as a function of radius for cool core ACCEPT groups and clusters (core entropy $\lesssim 30$ keV cm$^2$; \citealt{cav09}), and our models including the BCG potential. ACCEPT data is averaged in 5 temperature bins:  $<1$ keV ($<3.5 \times 10^{13} \msun$; grey filled circles); $1-2$ keV ($3.5 \times 10^{13}-1.1\times 10^{14} \msun$; red stars); $2-4$ keV  ($1.1-3.6 \times 10^{14}$; blue triangles); $4-6$ keV ($3.6-7 \times 10^{14} \msun$; green squares); $6-8$ keV ($7-11 \times 10^{14} \msun$; violet hexagons). The temperature range is converted into mass range using the best fit $M_{200}-T_X$ relation for our core entropy models in Table \ref{tab:tab4}; $M_{200} \approx 3\times 10^{14} \msun$ $( T_X /3.6~{\rm keV} )^{5/3}$. The theoretical profiles are shown for  halo masses of $10^{13},~6\times 10^{13},~3\times 10^{14},~6\times 10^{14},~10^{15} \msun$, in order of increasing density.}
\label{fig:nvsr1}}
\end{figure}
 
Figure \ref{fig:nvsr} shows the electron number density profiles for different halo masses using our constant entropy (solid lines) and constant $t_{\rm TI}/t_{\rm ff}$  (dotted lines) core models; the critical $t_{\rm TI}/t_{\rm ff}$ is 10 and the asymptotic gas density slope $s\equiv d\ln \rho/d\ln r=-2.25$. The halo mass ranges from that of the Galactic halo ($10^{12} \msun$) to a cluster ($3 \times 10^{14} \msun$). Figure \ref{fig:nvsr} shows several trends with the halo mass. Most importantly, the core electron density is smaller and the core size is larger for smaller mass halos. { }{Cool core clusters with mass $\gtrsim 10^{14} \msun$ have similar scaled density profiles according to our models, especially outside $0.05 r_{200}$. This is confirmed by observations of relaxed clusters with temperatures $\gtrsim 3.5$ keV (see Fig. 13 in \citealt{mau12}); density in smaller cool core halos is lower, in agreement with our models.} The constant $t_{\rm TI}/t_{\rm ff}$ cores give a much larger central density compared to the constant entropy cores, especially for smaller halos, { }{because $t_{\rm ff}$ decreases toward the center and the density has to increase to keep $t_{\rm TI}/t_{\rm ff}$ constant.}  

{ }{It has been observed that low mass halos (groups and elliptical galaxies) host cool `coronae'  instead of cool cores which are observed in massive clusters (e.g., \citealt{sun09}). These coronae have been likened to mini cooling flows which feed the central SMBHs and power radio jets. Instead of flat density cores, these coronae have density profiles steeply rising toward the center. We can explain the coronae in the context of our models if we include the gravitational potential of the central brightest cluster galaxy (BCG), in addition to the NFW potential. We model the gravitational acceleration due to the BCG as given by (in cgs units)
\be
g_{\rm BCG} = \left[  \left ( \frac{ (r/1 {\rm kpc})^{0.5975} }{3.206 \times 10^{-7}} \right )^s + \left ( \frac{ (r/1 {\rm kpc})^{1.849} }{1.861 \times 10^{-6}} \right )^s \right]^{-1/s}
\ee
 with $s=0.9$, and as shown in Figure 1 of  \citet{mat06}. We use the same BCG potential for all our halos. While this assumption is not correct in detail, it captures the essential feature that gravity at the smallest radii is dominated by the BCG and not by the NFW potential. }

{ }{ Figure \ref{fig:nvsr1} shows a comparison of our models (which include the BCG potential) with observations. The observations agree with our models in that the smaller halos have lower densities compared to self-similar modles. While we make the simplifying assumption that the gas density profile outside the `core' is the same for all halos, observations show that smaller halos have shallower density profiles. This is because realistic feedback can affect the gas even beyond the core, especially for the lower mass halos where the required feedback efficiency to balance cooling is quite small (see section \ref{sec:BB} and Fig. \ref{fig:epsvsM}). A comparison of Figures \ref{fig:nvsr} and \ref{fig:nvsr1} shows that the core density for $t_{\rm TI}/t_{\rm ff}=10$ models is much larger when the BCG potential is included. The effect is much more pronounced for smaller halos in which the BCG gravity dominates the NFW gravity even at relatively large radii. Thus, by including the BCG potential we are able to reproduce the observed steep rise in density toward the center in our low mass halo models.}

Most of our results stem from the fact that the smaller halos have lower core densities; in Section \ref{sec:LT} we study its implications for the $L_X-T_X$ relation, and in Section \ref{sec:BB} we apply it to the missing baryons problem.

 \subsection{$L_X-T_X$ Relation}
 \label{sec:LT}
 
Figure \ref{fig:nvsr} illustrates that the physics of thermal instability in halos results in bigger but lower density cores for smaller halos (cf. Fig. 9 in \citetalias{sha11}). This reduces the X-ray luminosity of low mass halos below the predictions of gravitational self-similar models. Moreover, Table \ref{tab:tab4} shows that the core entropy is nearly independent of the halo mass above $\gtrsim 3 \times 10^{13} \msun$, also in contrast with self-similar models. Thus, our models explain the observed steepening of the $L_X-T_X$ relation (\citealt{evr91}; \citealt{bry00} and references therein) and excess entropy (e.g., \citealt{pon99}) in lower mass halos.
 
 We can understand this result intuitively as follows. For self-similar evolution of hot gas in halos, we expect the gas density to be the same for all halo masses. This assumption leads to $L_X \propto n^2 \Lambda(T) r_{200}^3 \propto T^{3/2}\Lambda(T)$, which goes like $T^2$ for free-free cooling ($\Lambda \propto T^{1/2}$). On the other hand, holding  $t_{\rm TI} \propto T/n\Lambda(T)$ constant for all halos (assuming a similar $t_{\rm ff}$ for different halos) gives $n \propto T/\Lambda(T)$, which is $\propto T^{1/2}~{\rm or}~M_{200}^{1/3}$ for free-free cooling. Thus, our models predict lower densities for smaller halos, and hence $L_X \propto  n^2 \Lambda(T) r_{200}^3 \propto T^3$ for massive halos, in line with observations. Above arguments assume that most of the X-ray luminosity is contributed by the dense core and that the scaled core size is the same for all halos; as Figure \ref{fig:nvsr} shows, these assumptions are not satisfied quantitatively. The extra luminosity contributed by a bigger core for smaller halos is partly compensated by a smaller density because $t_{\rm ff}~({\rm roughly}~\propto r^{1/2})$ at larger radii is longer. Later we quantitatively discuss the $L_X-T_X$ relation obtained from our models.
 
 At halo temperatures below 1 keV, where line cooling becomes important and the cooling function starts to increase with a decreasing temperature, our ansatz of keeping  a fixed $t_{\rm TI}/t_{\rm ff}$ yields an even smaller core density and an even steeper $L_X-T_X$ relation. Observations suggest that such steepening does happen below 1 keV (e.g., \citealt{hel00}).
    
\begin{figure}
\centering
\includegraphics[scale=0.42]{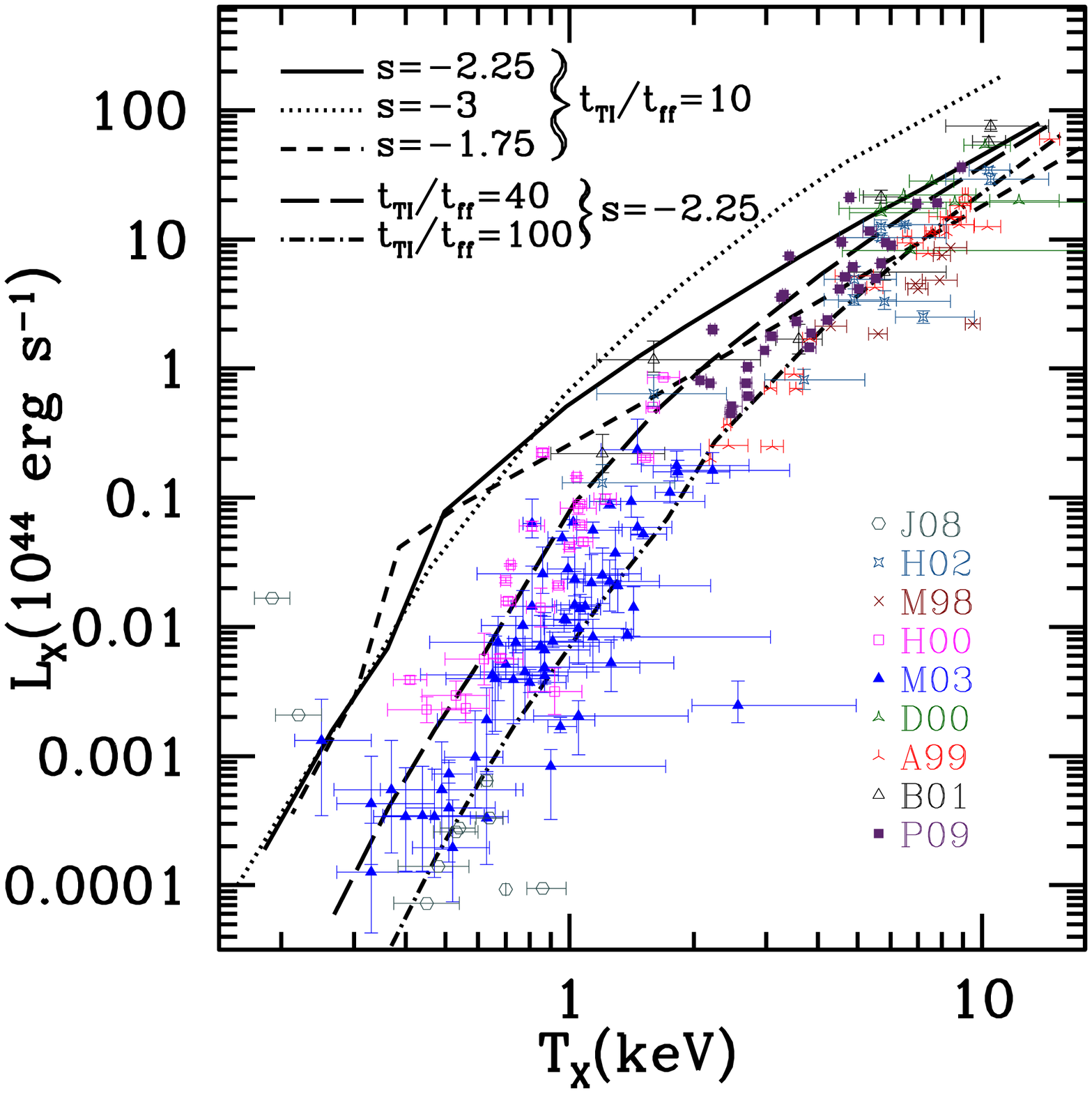}
\caption{The X-ray luminosity ($L_X$; Eq. \ref{eq:LX}) as a function of the luminosity-weighted temperature ($T_X$; Eq. \ref{eq:TX}) using our models and from observations. The observational data points are from \citealt{jel08}, \citealt{hol02}, \citealt{mar98}, \citealt{hel00}, \citealt{mul03}, \citealt{del00}, \citealt{arn99}, \citealt{bor01}, \citealt{pra09}, and references therein. The relations obtained from our one-dimensional models are: entropy core models with $t_{\rm TI}/t_{\rm ff} \geq 10$ and different outer gas density slopes ($s \equiv d\ln n/d\ln r$); entropy core models with $t_{\rm TI}/t_{\rm ff} \geq 40,~100$ and the outer density slope of $-2.25$. Increasing the $t_{\rm TI}/t_{\rm ff}$ threshold reduces the X-ray luminosity, especially for low mass halos.
\label{fig:LXvsTX}}
\end{figure}

More quantitatively, we directly  calculate the $L_X-T_X$ relation using the density and temperature profiles from our one-dimensional models (e.g., see Fig. \ref{fig:nvsr}). We calculate the luminosity-weighted temperature 
 \be
 \label{eq:TX}
 T_X=\frac{\int_0^{r_{200}} n_e n_i T \Lambda(T) 4\pi r^2 dr}{\int_0^{r_{200}} n_e n_i \Lambda(T) 4\pi r^2 dr},
 \ee
 and X-ray luminosity 
 \be
 \label{eq:LX}
 L_X=\int_0^{r_{200}} n_e n_i \Lambda(T) 4\pi r^2 dr.
 \ee
The solid line in Figure \ref{fig:LXvsTX} shows the $L_X-T_X$ relation for our fiducial model with an entropy core and $t_{\rm TI}/t_{\rm ff} \geq 10$ (corresponding to solid lines in Fig. \ref{fig:nvsr}); the asymptotic gas density slope is $-2.25$. The $L_X-T_X$ relation for the fiducial case provides a very good fit to the upper envelope of observational points. This is very satisfying because our $t_{\rm TI}/t_{\rm ff}=10$ model corresponds to an upper limit on the core density and hence on $L_X$; the core density and luminosity is lower than this for non-cool core halos. The models with a constant $t_{\rm TI}/t_{\rm ff}$ in the core, both with and without the BCG potential, give results similar to the entropy core models because the contribution of the increased density in the core is small. 
 
The long-dashed line and the dot-dashed line in Figure \ref{fig:LXvsTX} show the core entropy models with $t_{\rm TI}/t_{\rm ff}=40,~100$ respectively. For massive halos, the $L_X-T_X$ relation does not depend sensitively on the critical $t_{\rm TI}/t_{\rm ff}$ used for introducing the entropy core because the cores are small in these halos and contribute a subdominant fraction of the total luminosity. In lower mass halos, the core sizes are larger; thus, $L_X$ is more sensitive to the $t_{\rm TI}/t_{\rm ff}$ threshold. Any `overheating' by AGN or supernova feedback (which is reflected as an increase in $t_{\rm TI}/t_{\rm ff}$) could therefore contribute to the spread observed in the $L_X-T_X$ relation for lower mass halos.
 
Figure \ref{fig:LXvsTX} also shows that the outer gas density slope ($s \equiv d\ln\rho/d\ln r$) affects the $L_X-T_X$ relation for the high and low mass halos differently.  The luminosity is higher (lower) for clusters (groups) with steeper outer density profiles than for clusters (groups) with shallower profiles. For our lowest mass halos, where $t_{\rm TI}/t_{\rm ff} \lesssim 10$ even at $r_{200}$, the halo profiles are expected to be similar. The X-ray luminosity for a density profile with $d\ln \rho/d \ln r = -3$ is $\approx 4 $ times larger than for $d\ln \rho/d \ln r = -1.75$ at the massive halo end. Thus, the non-universality of the outer gas density profile~\citep[e.g.,][]{cro08}, { }{along with the variation in the core density}, can explain the large spread in $L_X-T_X$ relation for massive halos.

{ }{Recently \citet{mau12} have analyzed clusters with different temperatures and concluded that the relaxed hot ($\gtrsim 3.5$ keV) clusters have self-similar densities. Correspondingly, they find these clusters to satisfy the self-similar $L_X-T_X$ relation, $L_X \propto T_X^2$. The solid line in Figure \ref{fig:LXvsTX} shows that this is also the case for $T_X \gtrsim 1 $ keV clusters in our fiducial model corresponding to cool core halos. Self-similarity seems to persist until 1 keV in Figure \ref{fig:LXvsTX}, but in reality the density in smaller halos may be lower than given by Figure \ref{fig:nvsr} because of over-effectiveness of feedback in low-mass halos (Fig. \ref{fig:epsvsM} shows that the required feedback is smaller than AGN feedback efficiency for halos masses $\lesssim$ few$\times 10^{14} \msun$).}

An important fact that must be noted when comparing our models to observations is that the data in Figure \ref{fig:LXvsTX} are from different cluster/group/galaxy samples, both local and cosmological, and using different X-ray instruments. A lot of these observations are biased. For example, low temperature groups and galaxies have low surface brightness at large radii such that significant X-ray luminosity from close to the virial radius is missed (e.g., \citetalias{voi02}, \citealt{mul03}, \citealt{hel00}). Some of the data points have the contribution from the cool core removed (e.g., \citealt{mar98}), and some samples only contain clusters without cool cores (e.g., \citealt{arn99}).  The galaxies in \citet{jel08} are from groups and clusters, and are probably fainter than the field galaxies because of ram pressure and tidal stripping. Some data are from clusters at cosmological distances, thus its more difficult to detect plasma close to the virial radius (e.g., \citealt{del00,hol02}).  

While we calculate $L_X$ out to the virial radius $r_{200}$, most observations only go out to $r_{500}$. Recent observations show that the baryon content becomes closer to the universal value if the baryons are counted out to $r_{200}$ (e.g., \citealt{sim11,sat12}). Thus, for lower mass halos, where the density profile is shallower (see Fig. \ref{fig:nvsr}) and surface brightness is lower, large radii contribute significantly to $L_X$, and the observed luminosity is expected to be smaller than predictions. All these uncertainties preclude us from getting an exact match with observations. The overall trend of observations compared with predictions is very satisfying, however. \citet{cra10} argue that the $L_X-T_X$ relation for the X-ray halos of spiral galaxies is similar to the relation for ellipticals. This is expected if the X-ray halo is maintained in $t_{\rm cool}/t_{\rm ff} \gtrsim 10$ state in both spiral and elliptical galaxies via supernova and AGN feedback, respectively. 
  
 Previous works have tried to explain the low luminosity of galaxy groups relative to the scaling relations by expulsion of gas from smaller halos through AGN feedback (e.g., \citealt{puc08,mcc10,fab10}) or by mass dropout from the hot phase into the cold phase (e.g., \citealt{bry00,nag07}). However, there are problems with both scenarios. The mass dropout scenario without sufficient heating predicts massive cooling flows and excessive star formation, not observed even at group scales. The scenario where a large fraction of the X-ray emitting gas is blown out beyond the virial radius for groups, but not for clusters, is at odds with observations which show that the total halo baryon fraction within $r_{500}$, including the contribution of intracluster stars, is similar for clusters and groups (e.g., \citealt{gon07}). The gas fraction decreases with a decreasing halo mass, but the decrease in gas fraction is roughly compensated by the increase in stellar fraction. { }{By modeling stellar fraction in more detail, recent papers, e.g., \citealt{gio09,lea12}, show that there is a slight (less than a factor of 2) decline in the baryon fraction on going from massive groups to $10^{13} \msun$ halos. However, even for $10^{13} \msun$ halos the ICM mass is larger than stars, and a smaller X-ray surface brightness prevents an accurate measurement of the gas fraction, especially out to $r_{200}$.}
    
Our models roughly reproduce the observed density/entropy profiles (Fig. \ref{fig:nvsr}) and the observed  $L_X-T_X$ relation (Fig. \ref{fig:LXvsTX}). Moreover, since cooling (and probably feedback; see Section \ref{sec:BB})\footnote{Even if cooling is not important close to the virial radius, heating/outflows can, in principle, transport baryons out of the virial radius and thus reduce the baryon fraction (e.g., \citealt{puc08}).}  is not important close to the virial radius for halos $\gtrsim 10^{13} \msun$ (Fig. \ref{fig:nvsr}), we expect groups and clusters to be approximately closed systems (see $f_d$ in Fig. \ref{fig:epsvsM} and Table \ref{tab:tab4}), as the observations of \citet{gon07} suggest.
 
  \subsection{Baryon Budget in Halos}
 \label{sec:BB}
 
 \begin{figure}
\centering
\includegraphics[scale=0.42]{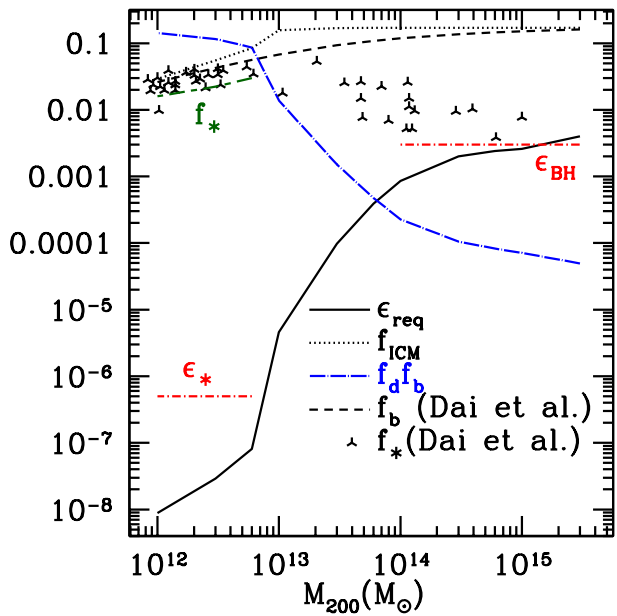}
\caption{Various quantities as a function of the halo mass for our fiducial model ({ }{results are not too different for $t_{\rm TI}/t_{\rm ff}=10$ `core' model}): the efficiency required to balance cooling ($\epsilon_{\rm req} \equiv L_X/\dot{M}_d c^2$; see Table \ref{tab:tab4} for the definition of various quantities);  the mass fraction expected to be present in the ICM, $f_{\rm ICM} \approx (1-f_d)f_b$; the dropout mass fraction, $f_df_b$; fit to the baryon mass fraction, $f_b$, using Eq. 7 from \citet{dai10}; and the stellar mass fraction, $f_\ast$, from Figure 5 of  \citealt{dai10}. Unlike them,  we use the halo mass $M_{200}$ as the independent variable, instead of the circular velocity, using $M_{200}=10^{15} \msun  [V_c/1632.07~{\rm km~s}^{-1}]^3$.  Also indicated are the typical ICM heating efficiencies expected due to AGN and supernovae ($\epsilon_{\rm BH}$ and $\epsilon_\ast$). Notice the sudden decrease in the required efficiency and the change in other quantities at $\sim 10^{13} \msun$. For halos $\lesssim 10^{13} \msun$ we expect most baryons to be beyond $r_{\rm 200}$ and only a small fraction ($f_\ast$ given by Eq. \ref{eq:fstar}; green long-short-dashed line) in form of stars.
\label{fig:epsvsM}}
\end{figure}

In this Section we use our models to study the baryon fraction as a function of halo mass. In our models the cluster mass halos tend to develop small cores (Fig. \ref{fig:nvsr}) and thus the dropout mass is insignificant; thus, clusters are baryonically closed with a baryon fraction similar to the cosmic value. Lower mass systems, however, can have a lower baryon fraction. In fact, for halo mass $\lesssim 10^{13} \msun$, $t_{\rm TI}/t_{\rm ff} \lesssim 10$ even at the outer boundary (see Fig. \ref{fig:nvsr}) and the ICM is depleted of most of the baryons, either due to star formation or because of expulsion of gas beyond the virial radius by feedback heating (see $f_d$ in Table \ref{tab:tab4}). Our models do not directly predict the fate of baryons absent from the ICM: whether they form stars or are lost due to feedback. However, the efficiency required to balance cooling ($\epsilon_{\rm req} \equiv L_X/\dot{M}_d c^2$; see Table \ref{tab:tab4}) as a function of halo mass provides some insight. If the required efficiency is small compared to the available feedback source, we expect the halo gas to be overheated, and sometimes even pushed beyond the virial radius. This is the case for halos less massive than $ 10^{13} \msun$. Thus, we expect these halos to have substantial baryonic mass `missing' from within $r_{200}$. This feedback-heated hot gas  beyond the virial radius is different from the warm-hot intergalactic medium (WHIM; \citealt{dav01}) which is heated by structure formation shocks in cosmological filaments. 

As argued above, a closed-box model (in which the inflow has the universal baryon fraction and outflows are negligible) for baryons is not good for halos less massive than $10^{13} \msun$. The equations of  mass and energy conservation in halos, in steady state, are
\ba
\label{eq:con_mass}
&& \dot{M}_{\rm in} \approx   \dot{M}_{\rm ICM}  + \dot{M}_{\rm BH} + \dot{M}_\ast + \dot{M}_{\rm out}, \\
\label{eq:con_energy}
&& \epsilon_\ast \dot{M}_\ast c^2 +  \epsilon_{\rm BH} \dot{M}_{\rm BH} c^2 \approx L_{X,{\rm ICM+out}} + \dot{E}_{\rm ICM} + \dot{E}_{\rm out},
\ea
where $\dot{M}_{\rm in}$ is the baryonic mass accretion rate onto the halo, $\dot{M}_{\rm ICM}$ is the rate of increase in the ICM mass, $\dot{M}_{\rm BH}$ is the mass accretion rate onto the central SMBH, $\dot{M}_\ast$ is the star formation rate, and $\dot{M}_{\rm out}$ is the mass outflow rate (outflows are driven by supernova and AGN feedback). Similarly, $\epsilon_\ast (\epsilon_{\rm BH}$) is the supernova (AGN) efficiency in heating the ICM, $L_{X,{\rm ICM+out}} $ is the X-ray luminosity of the ICM and the hot gas beyond $r_{200}$, $\dot{E}_{\rm ICM}$\footnote{Note that in absence of feedback heating and cooling, Eq. \ref{eq:con_energy} becomes $\dot{E}_{\rm ICM} \sim 0$, which reflects rough virial equilibrium. The weakly bound inflow virializes at the accretion shock and the gravitational energy is converted into thermal energy. Moreover, the timescale corresponding to $\dot{E}_{\rm ICM}$ is the halo age, which is much longer than the cooling time for groups and clusters. Detailed modeling of $\dot{E}_{\rm ICM}$ is beyond the scope of this paper.} is the rate of increase of the total ICM energy (internal + gravitational), and $\dot{E}_{\rm out}  \approx (5/2) f_{\rm vir} \dot{M}_{\rm out} k_B T_{200}/\mu m_p$ ($f_{\rm vir}$ is a factor comparing the temperature of the outflowing gas to $T_{200}$) is the total energy outflow rate from the halo. Eqs. \ref{eq:con_mass}-\ref{eq:con_energy} are severely underconstrained as stated, but observational data and our models can be used to constrain parameters such as $\dot{M_\ast}/\dot{M}_{\rm in}$, $\dot{M}_{\rm BH}/\dot{M}_{\rm in}$, $\dot{M}_{\rm out}/\dot{M}_{\rm in}$.  For example, the dropout fraction $f_d \equiv \dot{M}_d/\dot{M}_{\rm in} \approx 1- \dot{M}_{\rm ICM}/\dot{M}_{\rm in} $ can be read off from Table \ref{tab:tab4}. In massive halos ($\gtrsim 10^{13} \msun$ and $f_d \ll 1$), the dominant balance in mass conservation is $\dot{M}_{\rm in} \approx   \dot{M}_{\rm ICM} $; i.e., most of the mass accreted by the halos is incorporated in the ICM. Similarly, for massive halos Eq. \ref{eq:con_energy} reduces to $\epsilon_{\rm BH} \dot{M}_{\rm BH} c^2 \approx L_X$ because supernova feedback is expected to be subdominant, and $\dot{E}_{\rm ICM}$ and $\dot{E}_{\rm out}$ are negligible.\footnote{Note that a substantial contribution to $L_X$ comes from outer radii where the cooling time is long. Thus, AGN feedback need not power all of $L_X$; some of $L_X$ can come at the expense of $\dot{E}_{\rm ICM}$. Similarly, not all of the dropout mass is accreted by the SMBH. Here, we simply equate $L_X$ and $\epsilon_{\rm req} \dot{M}_d c^2$ to avoid introducing more free parameters.}

{ }{ Figure \ref{fig:epsvsM} shows the feedback efficiency required to balance cooling, $\epsilon_{\rm req} \equiv L_X/\dot{M}_d c^2$, as a function of the halo mass. The required efficiency decreases with a decreasing halo mass, as noted by \citet{gas11,gas12b}. }  The halos can be divided into three categories, depending on $\epsilon_{\rm req}$: first, the most massive halos ($\gtrsim  6 \times 10^{14} \msun$) in which $\epsilon_{\rm req}$ is larger than the efficiency of kinetic/radio-mode AGN feedback ($\epsilon_{\rm BH} \sim 0.003$;  e.g., \citealt{mer08}); second, the intermediate mass halos ($10^{13}-10^{14} \msun$) in which $\epsilon_{\rm req}$ lies between $\epsilon_{\rm BH}$ and $\epsilon_\ast~(\sim 10^{-6}-10^{-7}$);\footnote{We estimate the efficiency of supernova feedback as follows. We assume that the stars with $8 \msun \lesssim M_\ast \lesssim 20 \msun$ become supernovae (e.g., \citealt{woo02}) and assume a Salpeter IMF. The mass fraction in stars which will become supernovae is $\approx 0.07$. Only a small fraction ($\sim 0.1$) of the total supernova kinetic energy ($\sim 10^{51}$ erg) couples to the hot phase because of radiative losses (e.g., \citealt{tho98}). Thus, the supernova efficiency for heating the ICM  is $\sim 0.07\times0.1\times10^{51}/8{\rm M}_\odot c^2 \sim 5 \times 10^{-7}$ (indicated in Fig. \ref{fig:epsvsM}).} and third, the lowest mass halos in our study ($\lesssim 10^{13} \msun$) in which $\epsilon_{\rm req} \ll \epsilon_\ast$.

After introducing the three halo classes based on $\epsilon_{\rm req}$, we can study heating in each of these classes. The required efficiency for the most massive halos ($\sim 10^{15} \msun$) is $\gtrsim 0.003$ (see Table \ref{tab:tab4}), larger than typical estimates for radio mode AGN feedback ($\sim 0.003$), especially given the fact that only a small fraction of dropout mass may feed the SMBH. Thus, a substantial contribution to heating in massive halos is probably due to non-feedback processes such as thermal conduction (e.g., \citealt{guo08}). For intermediate mass halos ($10^{13}-10^{15} \msun$) AGN feedback is adequate to balance cooling  (Fig. \ref{fig:epsvsM}); however, supernova feedback is still insufficient. While the required feedback efficiency ($\epsilon_{\rm req}$) is smaller than $\epsilon_{\rm BH}$, the majority of dropout mass may lead to star formation instead of accretion. However, if a fixed fraction of $\dot{M}_d$ is channeled into $\dot{M}_{\rm BH}$ for all halos, galaxy groups are expected to be mostly in a non-cool core state  because a larger feedback efficiency results in a lower frequency of cooling episodes (see Fig. 11 in \citetalias{sha11}). For intermediate ($10^{13}-10^{15} \msun$) halos feedback heating is unlikely to expel baryons beyond the virial radius because the core is well within the virial radius (Fig. \ref{fig:nvsr}) and excess heating is expended in expansion of the core ($\dot{E}_{\rm ICM}$ in Eq. \ref{eq:con_energy}). Thus, even intermediate class halos are expected to be roughly baryonically closed.

The required efficiency ($\epsilon_{\rm req}$) is very small ($\lesssim 10^{-7}$) for the lowest mass halos ($\lesssim 10^{13} \msun$; see Fig. \ref{fig:epsvsM}). The feedback efficiency expected from star formation (mainly via supernovae) is $\sim 10^{-6}-10^{-7}$, smaller compared to the AGN feedback efficiency. But even this is  much larger than the efficiency required to power the X-ray luminosity in lower mass halos (see Fig. \ref{fig:epsvsM}). More quantitatively, a comparison of $\epsilon_{\rm req}$ and $\epsilon_\ast$ in Figure \ref{fig:epsvsM} shows that every baryon in the hot phase of smaller halos ($\lesssim 10^{13} \msun$) has $\sim 10-100$ times more energy available than is required to power its X-ray luminosity. Because of such a strong feedback heating, we expect  most  dropout baryons in lower mass halos to not form stars but instead be expelled beyond the virial radius. Moreover, the fraction of `missing' baryons in low mass halos is expected to increase with the decreasing halo mass because of a decreasing $\epsilon_{\rm req}$. 

Assume that the fraction of mass accreting onto the SMBH in low mass halos is negligible. Let us define $f_\ast \equiv f_b \dot{M}_\ast/\dot{M}_{\rm in}$, $f_{\rm out} \equiv f_b \dot{M}_{\rm out}/\dot{M}_{\rm in}$, $f_{\rm ICM} \equiv f_b \dot{M}_{\rm ICM}/\dot{M}_{\rm in}$, and $\epsilon_{\rm req} \equiv L_X/\dot{M}_d c^2$, where $\dot{M}_d = \dot{M}_\ast + \dot{M}_{\rm out}$. Thus, from Eq. \ref{eq:con_mass}, $f_b f_d \approx f_\ast + f_{\rm out}$ and $f_{\rm ICM} + f_{\rm out} + f_\ast \approx f_b$. The energy conservation equation (Eq. \ref{eq:con_energy}) becomes
\be
\label{eq:star_energy}
\epsilon_\ast \dot{M}_\ast c^2 \approx L_X \left (1 + \frac{f_{\rm out}}{f_{\rm ICM}} \right ) + \frac{5}{2}  \frac{k_BT_{200}}{\mu m_p }  f_{\rm vir}  \dot{M}_{\rm out},
\ee
where supernova feedback is assumed to power the X-ray luminosity of the ICM and the extended halo, and the thermal heating of the outflowing hot gas. Using our definitions, the above equation reduces to
\be
\label{eq:fstar}
\frac{f_\ast}{f_b} \approx 1 - \frac{\epsilon_\ast  +  (5/2)f_{\rm vir} (1-f_d) k_BT_{200}/\mu m_p c^2 }{ [\epsilon_\ast + \epsilon_{\rm req} f_d/(1-f_d) + (5/2) f_{\rm vir} k_BT_{200}/\mu m_p c^2] }, 
\ee
where $f_{\rm vir}$ is a virial factor comparing the temperature of the outflowing gas to the virial temperature ($T_{200}$).  { }{Note that the above equation is only valid when $\epsilon_\ast > \epsilon_{\rm req}$, i.e., only for halos $\lesssim 6 \times 10^{12} \msun$.} Eq. \ref{eq:fstar} should only be treated as an order of magnitude expression because the cooling time for low mass halos at large radii is long ($\sim 10$ Gyr) and our assumption of thermal balance need not hold as tightly. Moreover, the fraction of the supernova feedback energy going into heating the outflowing gas (the $f_{\rm vir}$ terms in Eqs. \ref{eq:star_energy} \& \ref{eq:fstar}) is very uncertain. The exact distribution of baryons in stars vs. `missing' baryons (beyond the virial radius) will require detailed numerical simulations beyond the scope of this paper. Here we only provide an order-of-magnitude estimate for the baryon budget in low mass halos. 

Figure \ref{fig:epsvsM} shows $f_\ast$ as a function of the halo mass for lower mass halos using Eq. \ref{eq:fstar} and setting $f_{\rm vir}$ to zero (green long-short-dashed line). The expectation is that the overestimated X-ray luminosity in Eq. \ref{eq:star_energy} (its an overestimate because the density beyond the virial radius is assumed to be the same as the core density) is of the order of the contribution of feedback in driving outflows. Plugging in $f_{\rm vir} \lesssim 1$ in Eq. \ref{eq:fstar} should not change our results significantly. The predicted stellar fraction for low mass halos shows a good match with observations. Figure \ref{fig:epsvsM} also indicates that excessive feedback in halos $\lesssim 10^{13} \msun$ results in thermal expulsion of majority of baryons beyond the virial radius ($f_{\rm out} \approx f_b - f_\ast -f_{\rm ICM} > f_\ast,~f_{\rm ICM}$). It is likely that these baryons were never incorporated within halos but were kept beyond $r_{200}$ because of excessive heating, as suggested observationally by \citet{and10}. 

Recall that we choose $f_b=0.17$, the universal value, in constructing our models (see Table \ref{tab:tab4}). However, as we argued in the previous paragraph, most baryons in low mass halos ($\lesssim 10^{13} \msun$) are likely to be expelled beyond $r_{200}$. Thus, the baryon fraction for lower mass halos, only accounting for baryons within the virial radius, should be much smaller than the universal value. Indeed, the total baryon fraction within $r_{200}$, $f_\ast + f_{\rm ICM}$, in Figure \ref{fig:epsvsM} is much smaller than 0.17 for halos less massive than $10^{13} \msun$. Table \ref{tab:tab4} shows that the mass dropout rate ($\dot{M}_d$) for the Milky Way  halo ($10^{12} \msun$) is { }{$\approx 28.2 \mpy$}, but our models suggest that the majority { }{($\sim 90\%$)} of these baryons are likely to be beyond $r_{200}$ { }{(Fig. \ref{fig:epsvsM})}. Thus, the star formation rate expected for the Galaxy is { }{$\sim 0.1 \dot{M}_d$}, which is $\sim 2-3 \mpy$ (see Table \ref{tab:tab4}), { }{quite similar} to the observed value (\citealt{bau10} and references therein).

Observationally, most of the baryons for halos $\gtrsim 10^{13} \msun$ are in the form of the ICM (see also \citealt{gio09}) and a closed box model for massive halos is a good approximation. Our prediction for the mass fraction in the ICM (dotted line in Fig. \ref{fig:epsvsM}) agrees qualitatively with the best-fit line from \citet{dai10}. In particular, the sharp decline in baryon fraction for halos $\lesssim 10^{13} \msun$ (corresponding to the temperature of $\approx$ 1 keV, as noted by \citealt{dai10}) is correctly captured by our models. Quantitative agreement is difficult because observational data show large scatter about the best fit. The observed mass fraction in stars also shows a large scatter and is much larger than the predicted dropout fraction for massive halos ($\gtrsim 10^{13} \msun$) because our models only consider accretion of baryons in form of gas. However, galaxies composed of stars, falling along cosmological filaments, are also accreted by massive halos. Thus, Figure \ref{fig:epsvsM} suggests that the stars from previous generation of galaxies (which merge to form the central BCG, and contribute to other galaxies and intracluster stars) form the majority of stars in galaxy clusters; the contribution of stars due to cooling of the hot ICM is negligible.

{ }{Galaxy clusters seem to pose an apparent paradox. The observed stellar mass fraction in clusters is smaller than in $10^{12}-10^{13} \msun$ halos (see Fig. \ref{fig:epsvsM}). However, clusters are formed via hierarchical mergers of star-rich smaller halos. How can the stellar fraction in clusters be smaller than in smaller halos? This paradox can be resolved by noting that the stellar fraction in even smaller ($\ll 10^{12} \msun$) halos falls precipitously with a decreasing halo mass (Fig. 5 in \citealt{dai10}). Thus, if a significant fraction of cluster growth occurs via accretion of very small halos (\citealt{fak10} suggest that smooth accretion contributes predominantly to the halo growth), clusters can have a smaller stellar fraction than $\sim 10^{12} \msun$ halos. As discussed in the previous paragraph, our models which only consider gas accretion obtain a stellar fraction much lower than in observed clusters. The detailed prediction of stellar fraction for clusters will require halo merger trees, and is beyond the scope of the present paper.}
 
\section{Conclusions}
 
 Inspired by our previous numerical simulations of the local thermal instability in the ICM with feedback heating, we have constructed simple one-dimensional models of  the ICM. According to our simulations, local thermal instability in the ICM leads to multiphase gas if, and only if, the ratio of the thermal instability timescale and the free-fall time ($t_{\rm TI}/t_{\rm ff}$) is $\lesssim 10$. For densities higher than this (i.e., $t_{\rm TI}/t_{\rm ff} \lesssim 10$) massive cold filaments will condense out of the hot phase, and the hot ICM will adjust roughly to a state with $t_{\rm TI}/t_{\rm ff} \gtrsim 10$ everywhere. Figure \ref{fig:nvsr} shows that the core density decreases with a decreasing halo mass; the decrease is quite pronounced for halos less massive than $10^{13} \msun$. Observations also show that the cores are bigger and more  tenuous in smaller mass halos (e.g., \citealt{sun12}). A lower core density for smaller halos naturally produces the steepening of the $L_X-T_X$ relation for lower mass halos (see Fig. \ref{fig:LXvsTX}). We do not expect the density profile, the $L_X-T_X$ relation, and the baryon budget to depend sensitively on redshift, as long as the feedback sources (AGN, supernovae) are active and the cooling time is shorter than the halo age. Thus, the assumption of a constant gas fraction for clusters at all redshifts, which is used to constrain cosmological parameters (e.g., \citealt{all08}), is expected to be valid.
  
 According to our models, the density of the Galactic halo is expected to be small  ($n_e \sim 10^{-4}$ cm$^{-3}$) because of a short  cooling time at the virial temperature of the Galactic halo ($\sim {\rm few} \times 10^{6}$ K). Thus most of the plasma in the ICM is expected to either cool and form stars or be expelled beyond the virial radius (see $f_d$ in Table \ref{tab:tab4}). Due to excessive feedback heating by supernovae, most of the baryons are pushed out of the virial radius instead of forming stars (see $f_\ast$ and $f_{\rm ICM}$ for halos $\lesssim 10^{13} \msun$ in Fig. \ref{fig:epsvsM}). Our models show that the halos less massive than $10^{13} \msun$ should be `missing' a substantial fraction of their baryons, in agreement with observations (e.g., \citealt{dai10}). 
 
 We tested two different models for the central cores, which make very different predictions for the gas density in the Galactic halo ($\sim 10^{12} \msun$). Observationally it has not been possible to unambiguously characterize the Galactic halo because of the low number density in the hot phase \citep[e.g.,][]{bre07}. However, pulsar dispersion measure and absorption line studies suggest that the number density is quite low ($n_e \sim 10^{-4}$ cm$^{-3}$; e.g., \citealt{and10}). A recent study which combines the Galactic absorption and emission observations of the hot gas points to a similar value (\citealt{gup12}). There are also indications (e.g., \citealt{eve08} and references therein) that some portions of the Galactic hot gas have densities as high as $\sim 0.01$ cm$^{-3}$, consistent with the prediction of our constant $t_{\rm TI}/t_{\rm ff}$ `core' models { }{which include the gravitational influence of the central galaxy.} 
 
Extended X-ray emission correlated with the indicators of star formation (UV, FIR, H$\alpha$) has been observed in halos of some spiral galaxies. The X-ray emission is believed to be powered by supernovae (e.g., \citealt{tul06,str04}) but some of it may also be due to the larger emissivity of a denser ICM. If so, this correlation of the low-entropy, high emissivity X-ray gas and the signatures of feedback (star formation in case of halos of spiral galaxies) is analogous to the correlation between the entropy and cold gas/AGN feedback in clusters of galaxies (e.g., \citealt{cav08}). Along similar lines, \citet{tum11} have recently observed large quantities of cooling gas (OVI) in outskirts of star forming galaxies, indicating a causal connection between the cooling of the hot halo and star formation. Our thermal instability + feedback models for hot halos try to explain the baryonic structure of all halos---from galaxies to clusters---in a single framework.
  
 According to our nonlinear criterion for the condensation of cold filaments in a gravitational field ($t_{\rm TI}/t_{\rm ff} \lesssim 10$) the high velocity clouds (HVCs) of atomic hydrogen in Galactic mass halos can condense out of the hot phase close to (or even beyond) the virial radius ($\sim 200$ kpc). This multiphase gas at distances large compared to the optical size of galaxies can explain quasar absorption lines observed in the neighborhood of galaxies (e.g., \citealt{che10}). A much larger density of the hot gas is required for the HVCs to condense out at smaller radii because of a short free-fall time (see the dotted line corresponding to the $t_{\rm TI}/t_{\rm ff}=10$ `core' model for the halo mass of $10^{12} \msun$ in Fig. \ref{fig:nvsr}). Once HVCs form, these dense, low metallicity filaments will fall freely toward the center of the halo, as observed (e.g., observations of M31 by \citealt{thi04}). These clouds will not only be shredded by Kelvin-Helmholz instabilities at the interface (e.g., \citealt{mur93}) but also reform due to efficient cooling of the shredded cool gas (see Fig. 4 in \citetalias{sha11}).
 
 We have intentionally kept our models simple by using the NFW gravitational potential for most halos; { }{we have also included a fixed BCG potential for some of the models (see Fig. \ref{fig:nvsr1}). 
The BCG models can explain the non-cool core clusters and groups which show much smaller but dense `coronae' that can power AGN feedback (e.g., \citealt{sun09}).} 
 Similarly, some isolated elliptical galaxies, in which the stellar bulge contributes significantly to the gravitational potential at inner radii and reduces the free-fall time, can have larger central densities (e.g., \citealt{wer12,hum12}).

From our models based on $t_{\rm TI}/t_{\rm ff}$, we expect the elliptical galaxies in cluster centers (the BCG, in particular) to have gas densities much    larger than similar mass field galaxies because the virial temperature, and hence the cooling time, is smaller for isolated galaxies. There is some observational support for this hypothesis (e.g., \citealt{mul10,sun09a}), but more observations are needed. We expect a lower ICM density for non-central elliptical galaxies in clusters, where the gravity of the dark matter halo does not point toward the center of the elliptical galaxy; ram pressure and tidal stripping introduce additional effects which further lower the hot gas density. Detailed environmental effects on the structure of hot gas in elliptical galaxies can only be studied with numerical simulations, and not by our simple models.
  
 \section*{Acknowledgments}

We thank Ian McCarthy, Massimo Gaspari, Biman Nath, Jerry Ostriker, Ming Sun, and especially James Bullock for useful discussions. We thank the anonymous referee for several constructive suggestions which improved the presentation. M. M., I. P. and E. Q. were supported in part by NASA Grant NNX10AC95G, NSF-DOE Grant PHY-0812811, Chandra Observatory theory grant TM2-13004X, and the David and Lucile Packard Foundation.

\onecolumn 
\begin{table}
\label{lastpage}
\caption{One Dimensional Models with an Entropy ($t_{\rm TI}/t_{\rm ff}$) Core using critical $t_{\rm TI}/t_{\rm ff} = 10$ and $s=-2.25$ 
\label{tab:tab4}}
\resizebox{\textwidth}{!}{%
\begin{tabular}{ccccccccc}
\hline
\hline
$M_{200}^\dag$ ($\msun$) & $c^{\ddag}$ & $n_{e,0}$ (cm$^{-3}$) & $K_0$ (keVcm$^2$) & $T_X$ (keV) & $L_X$ ($\times 10^{44}$erg s$^{-1}$) & $f_d^{\ddag\ddag}$ & $\dot{M}_d^\ast(\mpy)$ & $\epsilon_{\rm req}^{\dag\dag}$ \\
\hline
$3\times 10^{15}$ & 4  & 0.098 (0.15) & 20.7 (11.1) & 13.8 (13.7) & $79~(79.2)$ & $2.88~(0.37) \times 10^{-4}$ &29.3 (3.8) & 0.0047 (0.037) \\
$10^{15}$ & 5 & 0.071 (0.13) & 22.6 (8.68) & 7.07 (7.04) & $24.1~(24.2)$ & $4.14~(1.1) \times 10^{-4}$ & 14.1 (3.75) & 0.003 (0.013) \\
$6 \times 10^{14}$ & 6  & 0.069 (0.14) & 22.3 (7.7) & 5.4 (5.3) & $15~(15.2)$ &  $4.82~(1.84) \times 10^{-4}$ & 9.8 (3.75) & 0.0027 (0.0071) \\
$3 \times 10^{14}$ & 7 & 0.058 (0.13) & 22 (6.58) & 3.6 (3.59) & $7.3~(7.44)$ & $6.12~(3.8) \times 10^{-4}$ & 6.25 (3.9) & 0.0021 (0.0034) \\
$10^{14}$ & 8  & 0.038 (0.095) & 21.3 (5.3) & 1.91 (1.88) & $ 2.1~(2.17)$ & $1.32~(1.31) \times 10^{-3}$  & 4.5 (4.4) & 0.00082 (0.00085) \\
$6\times 10^{13}$ & 9 & 0.032 (0.088) & 21 (4.77) & 1.46 (1.44) & $1.23~(1.28)$ & $2.77~(2.37) \times 10^{-3}$ & 5.65 (4.84) & 0.00039 (0.00046) \\
$3\times 10^{13}$ & 9.5 & 0.02 (0.065) & 23.2 (4.34) & 0.98 (0.96) & $0.51~(0.55)$ & $8.8~(5.8) \times 10^{-3}$ & 8.99 (5.95) & $9.86 \times 10^{-5}~(0.00016)$ \\
$10^{13}$ & 10.5 & 0.0045 (0.035) & 47.9 (4.04) & 0.5 (0.51) & $0.078~(0.13)$ & 0.08 (0.036) & 27.3 (12.1) & $5 \times 10^{-6}~(1.88\times10^{-5})$ \\
$6 \times 10^{12}$ & 11 &  $6.4 \times 10^{-4}$ (0.023) & 159.6 (4.11) & 0.36 (0.38) & $6.79\times 10^{-3}~(0.045)$ & 0.5 (0.2) & 102.6 (41.3) & $1.15 \times 10^{-7}~(1.92 \times 10^{-6})$\\
$3 \times 10^{12}$ & 11.5 & $3.3 \times 10^{-4}$ (0.011) & 171.1 (4.56) & 0.27 (0.26) & $1.58~(7.7) \times 10^{-3}$ & 0.68 (0.52) & 68.9 (53.4) & $4 \times 10^{-8}~ (2.5\times 10^{-7})$\\
$10^{12}$ & 12 & $1.2 \times 10^{-4}$ ($3 \times 10^{-3}$) & 193.3 (6.16) & 0.18 (0.16) & $1.88~(5.69) \times 10^{-4}$ & 0.83 (0.78) & 28.2~(26.4) & $1.2 \times 10^{-8}~(3.8\times 10^{-8})$  \\
\hline
\end{tabular}}
{The halo baryon mass fraction is chosen to be the universal baryon fraction $f_b=0.17$ for all halos. The entropy core model is the fiducial model. The quantities in parentheses denote the values for a constant $t_{\rm TI}/t_{\rm ff}$ `core' model. The various quantities are: $n_{e,0}$ the central electron density (at $r=0.001r_{200}$), $K_0$ the central entropy, $T_X$ the luminosity-weighted temperature (Eq. \ref{eq:TX}), and $L_X$ the X-ray luminosity (Eq. \ref{eq:LX}).} \\
{$^\dag$}{$M_{200}$ is the halo mass within $r_{200}$, such that, $M_{200}  = (4\pi/3) 200 \rho_{\rm cr} r_{\rm 200}^3$, where $\rho_{\rm cr}$ is the critical density taken to be $9.2 \times 10^{-30}$ g cm$^{-3}$.} \\
{$^\ddag$}{$c$ is the concentration parameter defined as $c=r_{200}/r_s$, where $r_s$ is the scale radius.} \\
{$^{\ddag\ddag}$}{$f_d\equiv (f_bM_{200} - \int_0^{r_{200}} 4\pi r^2 \rho dr)/f_bM_{200}$ is the mass dropout fraction, where $\rho$ is the gas mass density.} \\
{$^\ast$}{$\dot{M}_d \equiv f_b f_d M_{200}/t_{\rm age} \approx \dot{M}_{\rm out} + \dot{M}_\ast + \dot{M}_{\rm BH} \approx \dot{M}_{\rm in} - \dot{M}_{\rm ICM}$ (see Eq. \ref{eq:con_mass}) is the mass dropout rate, where we choose $t_{\rm age} = 5$ Gyr.}\\
{$^{\dag\dag}$}{$\epsilon_{\rm req} \equiv L_X/\dot{M_d}c^2$ is the heating efficiency required to balance radiative cooling of the ICM.}
\end{table}

\end{document}